\documentclass[prb,twocolumn,showpacs,amsmath,amssymb,superscriptaddress]{revtex4-1}
\usepackage{dcolumn}
\usepackage{bm,graphicx}
\usepackage{etoolbox}
\usepackage{url}
\usepackage{hyperref}

\begin{document}

\title{Non-uniform carrier density in Cd$_3$As$_2$ evidenced by optical spectroscopy}

\author{I.~Crassee}
\affiliation{LNCMI, CNRS-UGA-UPS-INSA, 25, avenue des Martyrs, F-38042 Grenoble, France}

\author{E.~Martino}
\affiliation{IPHYS, EPFL, CH-1015 Ecublens, Switzerland}

\author{C.~C.~Homes}
\affiliation{Condensed Matter Physics and
   Materials Science Department, Brookhaven National Laboratory, Upton,
   New York 11973, USA}

\author{O.~Caha}
\affiliation{CEITEC and Masaryk University, Faculty of Science, CZ-61137 Brno, Czech Republic}

\author{J.~Nov\'ak}
\affiliation{CEITEC and Masaryk University, Faculty of Science, CZ-61137 Brno, Czech Republic}

\author{P.~T\"uckmantel}
\affiliation{DQMP, University of Geneva, CH-1211 Geneva 4, Switzerland}

\author{M.~Hakl}
\affiliation{LNCMI, CNRS-UGA-UPS-INSA, 25, avenue des Martyrs, F-38042 Grenoble, France}

\author{A.~Nateprov}
\affiliation{Institute of Applied Physics, Academy of Sciences of Moldova, MD-2028 Chisinau, Moldova}

\author{E.~Arushanov}
\affiliation{Institute of Applied Physics, Academy of Sciences of Moldova, MD-2028 Chisinau, Moldova}

\author{Q.~D.~Gibson}
\affiliation{Department of Chemistry, Princeton University, Princeton, New Jersey 08544, USA}

\author{R.~J.~Cava}
\affiliation{Department of Chemistry, Princeton University, Princeton, New Jersey 08544, USA}

\author{S.~M.~Koohpayeh}
\affiliation{Institute for Quantum Matter and Department of Physics and Astronomy,
The Johns Hopkins University, Baltimore, Maryland US-21218, USA}

\author{K.~E.~Arpino}
\affiliation{Institute for Quantum Matter and Department of Physics and Astronomy,
The Johns Hopkins University, Baltimore, Maryland US-21218, USA}
\affiliation{Department of Chemistry, The Johns Hopkins University, Baltimore,
Maryland US-21218, USA}

\author{T.~M.~McQueen}
\affiliation{Institute for Quantum Matter and Department of Physics and Astronomy,
The Johns Hopkins University, Baltimore, Maryland US-21218, USA}
\affiliation{Department of Chemistry, The Johns Hopkins University, Baltimore,
Maryland US-21218, USA}
\affiliation{Department of Materials Science and Engineering, The Johns Hopkins
University, Baltimore, Maryland US-21218, USA}

\author{M.~Orlita}
\affiliation{LNCMI, CNRS-UGA-UPS-INSA, 25, avenue des Martyrs, F-38042 Grenoble, France}
\affiliation{Institute of Physics, Charles University in Prague, CZ-12116 Prague, Czech Republic}

\author{Ana Akrap}\email{ana.akrap@unige.ch}
\affiliation{DQMP, University of Geneva, CH-1211 Geneva 4, Switzerland}

\date{\today}

\begin{abstract}
We report the detailed optical properties of Cd$_3$As$_2$ crystals in a wide parameter space: temperature, magnetic field, carrier concentration and crystal orientation. We investigate high-quality crystals synthesized by three different techniques. In all the studied  samples, independently of how they were prepared and how they were treated before the optical experiments, our data indicate conspicuous fluctuations in the carrier density (up to 30\%). These charge puddles have a characteristic scale of 100~$\mu$m, they become more pronounced at low temperatures, and possibly, they become enhanced by the presence of crystal twinning.
The Drude response is characterized by very small scattering rates ($\sim 1$~meV) for as-grown samples.  Mechanical treatment, such as cutting or polishing, influences the optical properties of single crystals, by increasing the Drude scattering rate and also modifying the high frequency optical response. Magneto-reflectivity and Kerr rotation are consistent with electron-like charge carriers and a spatially non-uniform carrier density.
\end{abstract}

\maketitle

\section{Introduction}

Cadmium arsenide Cd$_3$As$_2$ is a compound known and studied since the 1960's for its unusually high electron mobility. Recently it was proposed that Cd$_3$As$_2$ could be the realization of a three-dimensional Dirac semimetal.\cite{WangPRB13} Experiments such as ARPES, STM, and electrical transport measurements in high magnetic field soon confirmed the linear dispersion of electronic bands in a wide energy range.\cite{BorisenkoPRL14,Neupane2014aa,Liu2014aa,JeonNatureMater14} However, the experimentally observed linear dispersion appears not to be linked to any topological or symmetry properties of this compound. Instead, magneto-optical spectroscopy in high magnetic fields shows that Dirac cones are not experimentally accessible in Cd$_3$As$_2$ and a single Kane cone extends over a wide energy range.\cite{Akrap2016} While the Dirac cones are mandated by the symmetry of the compound, they are expected to be limited to very low energies around the Dirac points, less than 30~meV. At the same time, the Fermi level, determined with respect to the Dirac points, is between 80 and 200 meV in most of the available crystals.
The Kane model, applied to a system with a nearly vanishing band gap, successfully explains the existence of largely extended conical band.\cite{Orlita2014,Bodnar77,Akrap2016} Nevertheless, within its great simplicity, the model neglects all effects related to possible lack of inversion symmetry of Cd$_3$As$_2$. Hence, it cannot account for the complex response observed in magnetic oscillation measurements on crystals of Cd$_3$As$_2$, where two Fermi surfaces are observed.\cite{PariariPRB15, ZhaoPRX15, Moll2016}

\begin{figure}[t]
\includegraphics[trim = 0mm 0mm 0mm 0mm, clip=true, width=1\columnwidth]{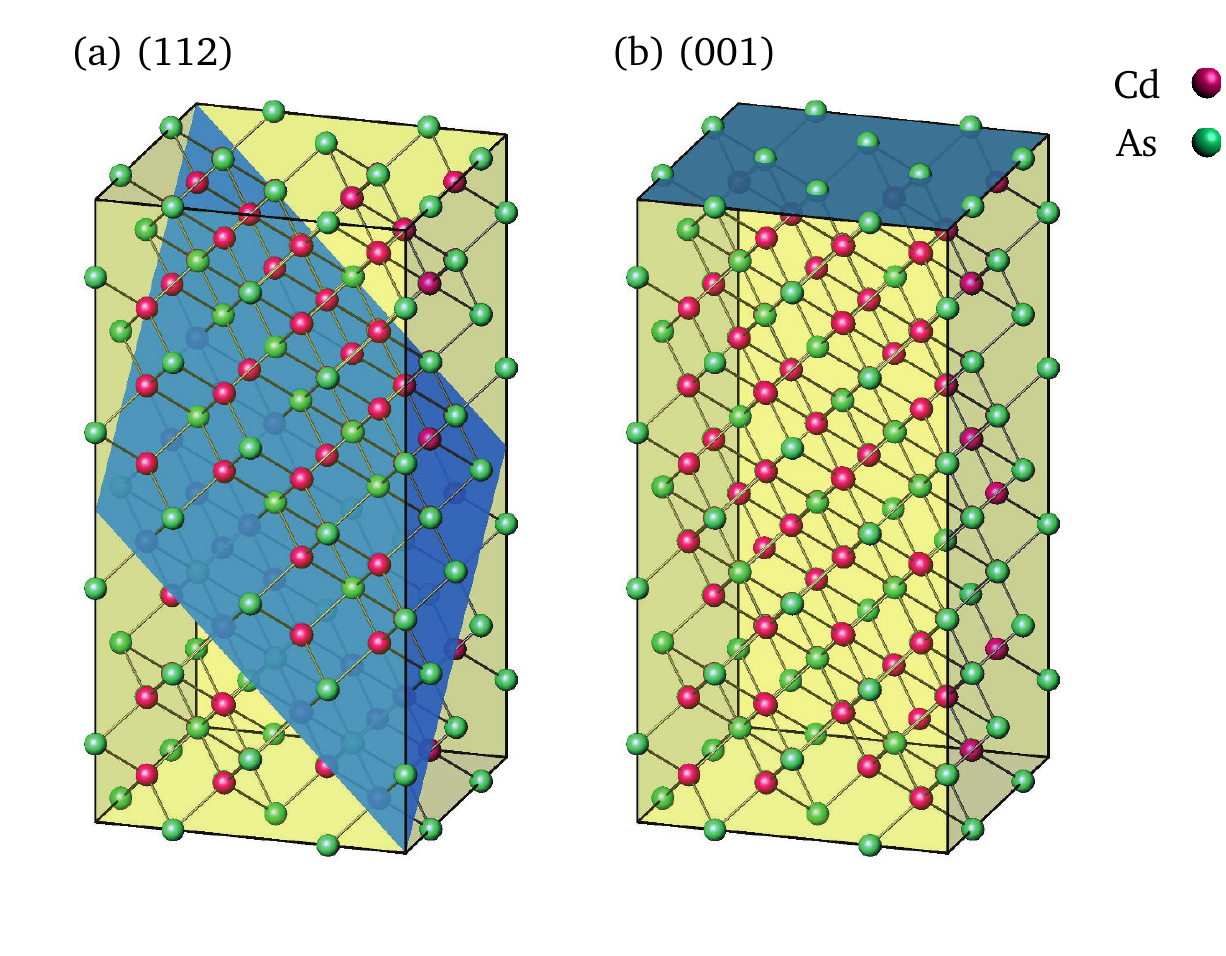}
\caption{\label{Fig0:Structure} (color online) The crystal structure of Cd$_3$As$_2$ is shown, highlighting the two planes we study: in (a), the naturally exposed (112) plane with a hexagonal motif; in (b), the optically isotropic (001) plane, perpendicular to the tetragonal axis.}
\end{figure}

It is important to take into account the remarkably complex material aspects of Cd$_3$As$_2$, which may dictate an intricate behavior of conducting electrons.
There are several polymorphic modifications of this compound. As the temperature decreases, the high-temperature fcc $\beta$-phase is followed by the lower symmetry $\alpha''$, $\alpha'$, and ultimately the $\alpha$-phase. Throughout all these phases, arsenic atoms form a slightly distorted close-packed cube. Cadmium atoms are placed in the tetrahedral voids, and two out of four voids are occupied. Cadmium atoms randomly fill the tetrahedral voids in the $\beta$-phase, but are highly ordered in the each of the $\alpha$ phases. As the temperature drops, each structural phase transition is accompanied by a sudden change in the lattice constants which then leads to micro-cracks in the crystal.\cite{Arushanov1986aa}
Ultimately the room-temperature $\alpha$-phase is given by a unit cell comprising 160 atoms. The unit cell is tetragonal but remains close to a cubic cell, with an antifluorite-like structure. Systematic cadmium vacancies are arranged such that they lead to a $C_4$ rotational  symmetry around the [001] axis, protecting the Dirac nodes.
Cadmium vacancies have been known to anneal at room temperature with time, leading to a decrease in the overall carrier density.\cite{Arushanov1986aa, Blom1979}

Morphologically, Cd$_3$As$_2$ crystals are bound by (112) faces, as illustrated in Fig.~\ref{Fig0:Structure}(a). The surfaces typically have a large number of steps. The (112) plane is optically anisotropic and contains no principal optical axes. The optically isotropic plane is the (001) plane, Fig.~\ref{Fig0:Structure}(b), which can be exposed by cutting and polishing the crystals. However, cadmium arsenide is mechanically soft and while the exact mechanisms are unknown, manipulations like thermal cycling, cutting or polishing the sample may damage the crystals and subsequently influence low-energy properties.

The aim of this paper is to shed light on the low-energy properties of Cd$_3$As$_2$, by means of infrared spectroscopy.
We study the detailed optical properties of differently synthesized (112) and (001)-oriented Cd$_3$As$_2$ crystals, for a wide range of parameters. These include carrier density, temperature, high pressure, magnetic field and mechanical polishing on the dynamical conductivity of Cd$_3$As$_2$. We find signatures of twinning and optical anisotropy.
Finally, we show that electron puddles are characteristic of bulk Cd$_3$As$_2$. Carrier density varies by 30\% on a scale of $\sim 100$~$\mu$m. Spatial non-uniformity occurs in all the samples investigated here, and it is expected to impact a number of electrical, optical and magneto-transport properties.

\begin{figure*}[t]
\includegraphics[trim = 0mm 0mm 0mm 0mm, clip=true, width=17cm]{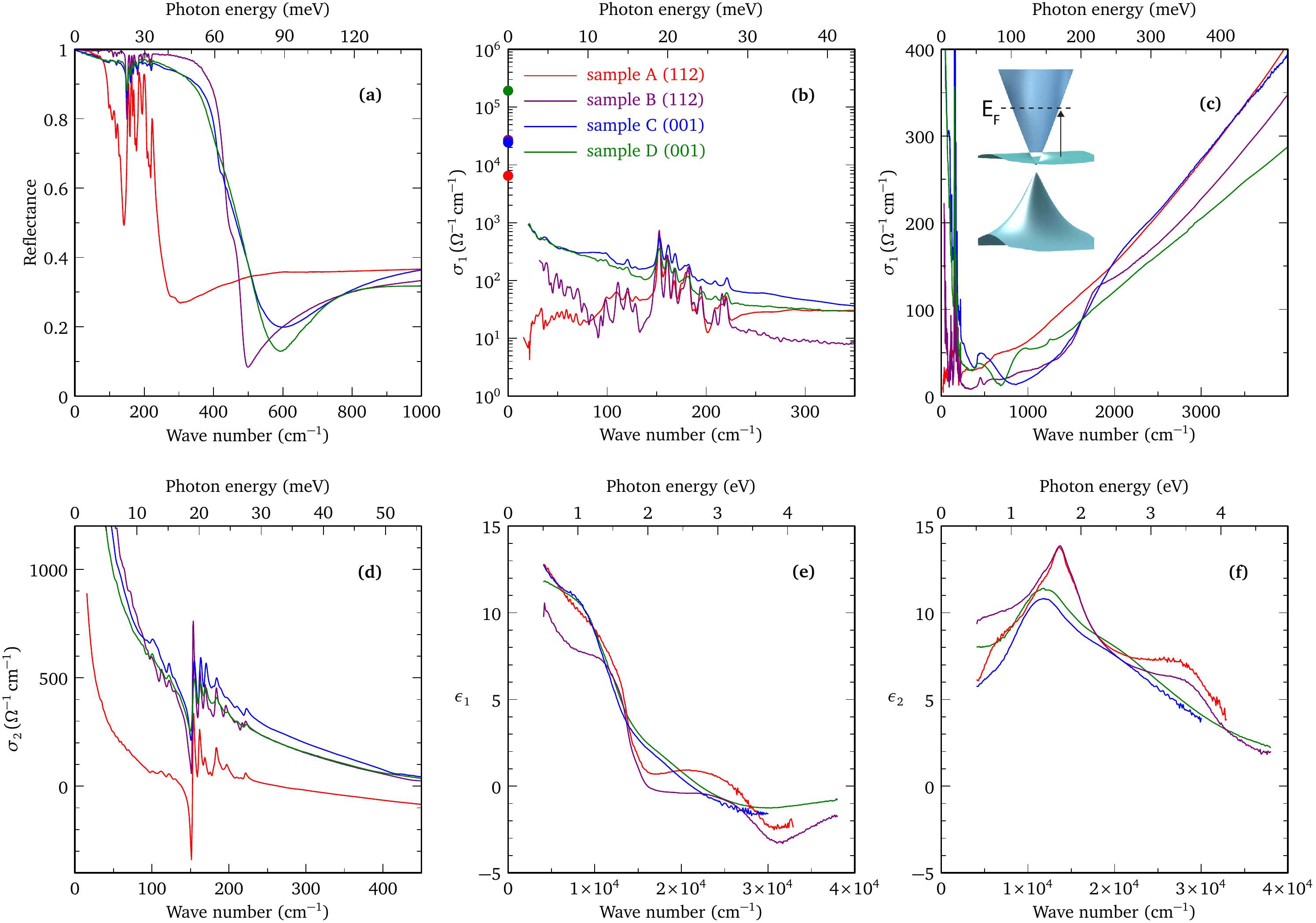}
\caption{\label{Fig1:Doping} (color online) (a) The reflectance of four studied samples is shown at 5~K and $B=0$, dominated by sharp plasma edges and phonon response. (b) and (c) Real part of the optical conductivity, $\sigma_1$, at 5~K. In (b), solid circles at zero photon energy indicate the values of $\sigma_{dc}$; maroon and blue circles overlap.
Inset in (c) shows the band structure determined from magneto-optical experiments.\cite{Akrap2016} The Fermi level is indicated, and the arrow shows the onset of interband absorption, seen as a step-like feature in $\sigma_1$.
(d) Imaginary part of the optical conductivity, $\sigma_2$, at 5~K. (e) Real and (f) imaginary parts of the dielectric function measured at room temperature in a broad range.}
\end{figure*}

\section{Experimental methods}

Experiments were performed on several $n$-doped single crystals of Cd$_3$As$_2$. Samples were grown employing three different techniques.
The (112)-oriented samples A and A$^\prime$ ($\sim$2~mm$\times$4~mm) were grown using the thermal evaporation method, as described in Refs.~\onlinecite{Arushanov1986aa} and \onlinecite{WeberAPL15}. Samples B and C ($\sim$1.5~mm$\times$1.5~mm) were grown out of Cd-rich flux as explained in Ref.~\onlinecite{AliIC14}. Sample D ($\sim$10~mm$\times$5~mm) was prepared using a traveling solvent technique, as detailed in Ref.~\onlinecite{Koohpayeh2016aa}.
For samples A and B, their natural (112) surfaces were investigated. Samples C and D were first oriented, then cut and polished in order to expose their optically isotropic (001) plane.

Optical experiments were done in reflection geometry using a FTIR spectrometer. Spectra were obtained for a set of different temperatures in the optical range from 3 meV (24~cm$^{-1}$, infrared) to 3~eV (24000~cm$^{-1}$, UV) using in situ gold evaporation for referencing.\cite{Homes1993} In addition, room temperature ellipsometry was employed to determine the dielectric function from 0.5 (4000~cm$^{-1}$) to 4~eV (32000~cm$^{-1}$).
Optical microscopy in the infrared was used to inspect the local variations of the low energy dynamics. Utilizing an aperture of approximately 100~$\mu$m$\times$100~$\mu$m in size, the reflectance of each sample was mapped across the entire surface of the crystal and referenced to a gold patch evaporated onto the sample in advance. In addition polarized light was used to study the local optical anisotropy.

Magneto-optical Kerr effect and magneto-reflectivity were measured on sample D with the exposed (001) facet, up to 7~T, using the methods described in Ref.~\onlinecite{Levallois2015}.

All the samples were characterized using X-ray scattering in both symmetric and asymmetric configuration, allowing us to check for crystal orientation and for possible twinning.
The x-ray diffraction was measured using a Rigaku SmartLab diffractometer, equipped with a copper x-ray tube and a two-bounce Ge(220) monochromator.
A scintillation counter was used as a detector.
In order to determine the surface orientation, we have measured a symmetric $\theta -2\theta$ scan. The azimuthal orientation and presence of twins were checked by performing an azimuthal scan for the 325 diffraction peak, which is far enough from any other diffraction peaks to avoid possible overlap with other diffractions.

\section{Results and discussion}

This section begins with a study of optical properties at various electron densities. The following part addresses the temperature dependence of reflectivity and dynamical conductivity, showing the low-temperature development of Pauli blocking edge and fine structure of the plasma edge.
Then we focus on position-dependent optical reflectivity, which directly shows the existence of charge puddles in Cd$_3$As$_2$. Finally, we discuss the magneto-reflectivity and the magneto-optical Kerr effect.

\subsection{Dependence of infrared properties on carrier density and temperature}

Optical properties are commonly described using the complex dielectric function: $$\tilde{\epsilon}(\omega)=\epsilon_1(\omega) + i \epsilon_2(\omega)$$
where $\omega$ is the incident photon frequency. Optical (dynamical) conductivity can then be determined in the following way:
$$\tilde{\sigma}(\omega)=-2\pi i \omega [\tilde{\epsilon}(\omega)-\epsilon_\infty]/Z_0 =\sigma_1(\omega)+ i \sigma_2(\omega)$$
where $\epsilon_\infty$ is the real part of the dielectric function at high frequency, $Z_0\approx 377\, \Omega$ is the impedance of free space, and the resulting optical conductivity $\sigma$ is in the units of $\Omega^{-1}$cm$^{-1}$.

\begin{figure*}[t]
\includegraphics[trim = 0mm 0mm 0mm 0mm, clip=true, width=16cm]{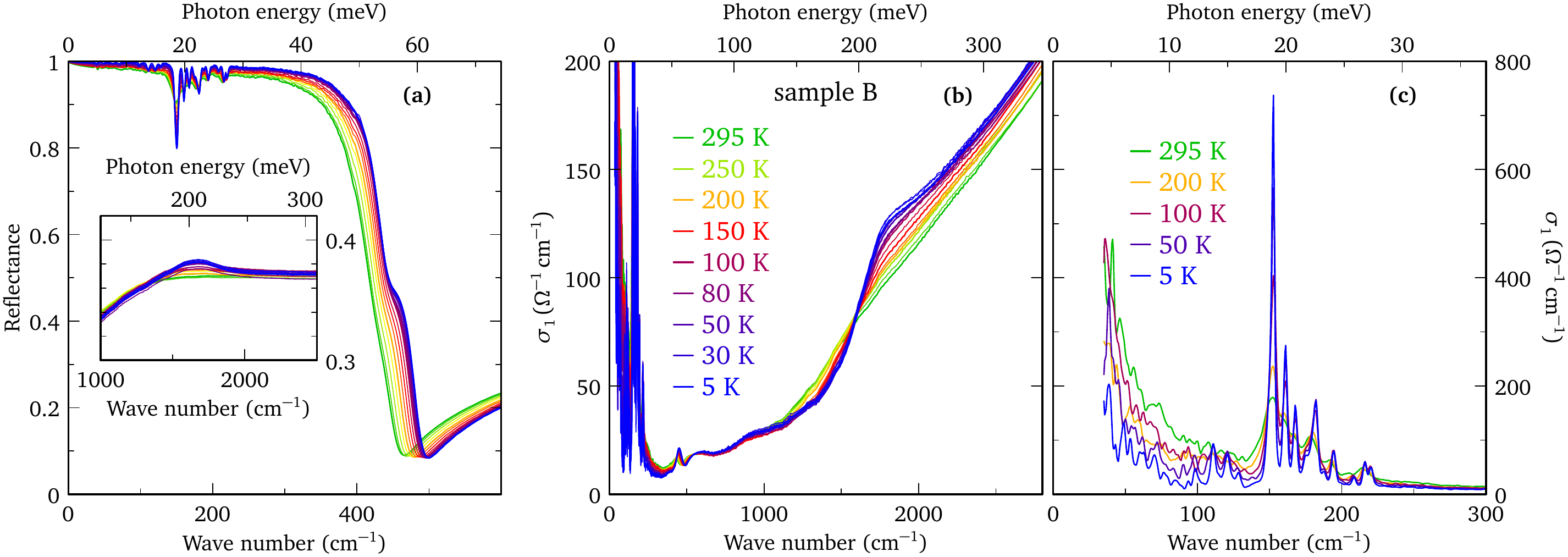}
\caption{\label{Fig2:Temperature} (color online) (a) The reflectance of sample B as a function of temperature, for low photon energies. Inset shows the feature in reflectance which corresponds to the Pauli blocking or the edge of interband absorption.  (b) and (c)  Real part of the optical conductivity, $\sigma_1$, as a function of temperature. Development of a step feature in (b) at $\sim 200$~meV is linked to the onset of interband absorption and Pauli blocking, as illustrated in the inset of Fig.~\ref{Fig1:Doping}(c).}
\end{figure*}

Figs.~\ref{Fig1:Doping}(a)--(d) shows the optical reflectance, as well as the real ($\sigma_1$) and imaginary ($\sigma_2$) parts of the complex optical conductivity, determined by Kramers-Kronig analysis, for four different samples of Cd$_3$As$_2$ at $T=5$~K. Samples A and B were measured on their as-grown mirror-like (112) surfaces, whereas samples C and D were oriented, cut and polished to expose the optically isotropic (001) planes.

The reflectance shows a sharp plasma edge deep in the far infrared for all the four samples, as illustrated in Fig.~\ref{Fig1:Doping}(a). The two (112)-oriented samples (A and B) have particularly sharp plasma edges, indicating a tiny Drude scattering rate.
All four samples are metallic-like at low frequencies, their reflectance approaching unity in the dc limit.
Rather low plasma frequencies indicate relatively small electron concentrations, lying in the range $2.5\times 10^{17} \sim 2.5\times 10^{18}$~cm$^{-3}$, as discussed later on.
The carrier concentration varies from one sample to another, and can be assessed from the plasma frequency:
$$ \omega_p^2 = \frac{n e^2}{\epsilon_0 m^*},
$$
where $n$ is the carrier density, and $m^*$ the effective electronic mass.
The screened plasma edge $\tilde{\omega}_p=\omega_p /\sqrt{\epsilon_\infty}$ may be taken as the position of the step in reflectivity, and it ranges from 200 to 500 cm$^{-1}$ (30 to 75~meV). The plasma frequency is directly proportional to the Fermi energy in a system with linear energy dispersion, so that the spread of  $\omega_p$ is reflective of a difference in Fermi level amongst our samples.
The lowest $\omega_p$ is found in sample A, while samples B, C and D have a comparable $\omega_p$. In the static limit, $\sigma_{dc}$ was determined by the electric transport measurements. $\sigma_{dc}$ was the smallest in sample A and the highest in sample D, as indicated in Fig.~\ref{Fig1:Doping}(b). Incidentally, sample A had been annealed at room temperature for the longest period of time, over several years.

The reflectance of each sample has a small notch or a doubled structure at the plasma edge. This is more evident in samples A and B which expose the as-grown surface (112), but it is visible as a change in the slope of the plasma edge in the two (001) samples at 525 cm$^{-1}$ (65~meV).
Very generally, this doubled structure of the plasma edge is indicative of more than a single contribution to the plasma excitation. For example, it may be due to a spatial inhomogeneity of carrier density, or due to the optical anisotropy.
In the two (001) samples, the plasma edge does not have a simple shape, and cannot be modelled by a single Drude contribution. This is possibly linked to the mechanical treatment (polishing).

Cd$_3$As$_2$ has a tendency to be $n$-doped due to an inherent arsenic deficiency.
There can be a huge variability of dc resistivity even within one batch of samples,\cite{Liang2015} wherein the low temperature resistivity was found to vary from 20~n$\Omega$cm to 400~m$\Omega$cm.

Figure~\ref{Fig1:Doping}(b) shows the real part of the optical conductivity $\sigma_1(\omega)$ in the far infrared region, on a logarithmic scale. At lowest frequencies $\sigma_1$ is dominated by a set of sharp phonon modes.
In the two (112) oriented samples, exposing a naturally grown plane, the modes are sharper than in the polished (001) samples. This broadening of the phonon lines may be linked to mechanical damage from the polishing.
The crystal structure is described by a centrosymmetric group $I4_1 acd$,\cite{AliIC14} and the factor group analysis gives the following allowed phonon modes:
\begin{multline*}
\Gamma = 14 A_{1g} + 15 A_{2g} + 14 B_{1g} + 15 B_{2g} + 31 E_{1g} \\
+ 14 A_{1u} + 14 A_{2u} + 14 B_{1u} + 15 B_{2u} + 30 E_{u}.
\end{multline*}
Among those, only the $A_{2u}$ and $E_u$ modes are infrared active phonon modes, giving in total 44 predicted phonon modes. $A_{2u}$ phonons are active along $c$-axis, and $E_u$ modes are active in the {\it a-b} plane.
Only 13 infrared active phonon modes can be seen from the experimental spectra. Four of those modes are situated below 150~cm$^{-1}$, namely at 98, 111, 120 and 129~cm$^{-1}$. Those modes are observed in the (112) plane, but not in the (001) plane, so we assign them to the $A_{2u}$ symmetry.
The remaining nine phonons are at 152.5, 160.7, 168, 176.7, 182.5, 195, 209, 215 and 220~cm$^{-1}$. These modes are observed for both sample orientations, and we assign them to the $E_u$ symmetry.
According to a recent Raman study which identified 44 Raman active modes, the higher frequency modes (above 150~cm$^{-1}$) are linked to As phonons, and the lower frequency modes correspond to the collective vibrations of mainly Cd displacements.\cite{Sharafeev2017aa} The infrared phonon modes are of different symmetry than the Raman modes in a centrosymmetric structure of Cd$_3$As$_2$. Yet, all of the infrared phonons observed here are very close in frequency (or accidentally degenerate) to the observed Raman-active phonons.\cite{Sharafeev2017aa} This is not surprising considering the large and complex unit cell.

A comparison between $\sigma_1(\omega)$ and $\sigma_{dc}$, shown in Fig.~\ref{Fig1:Doping}(b), implies that the Drude contribution is extremely narrow. While the values of $\sigma_{dc}$ span from 6500 to 190 000~$\Omega^{-1}$~cm$^{-1}$, they are orders of magnitude higher than the values of $\sigma_1$ in the far infrared, which range from 10 to 1000~$\Omega^{-1}$~cm$^{-1}$.
In fact, in sample A, the Drude contribution cannot even be directly seen in $\sigma_1$ down to 30~cm$^{-1}$ (4~meV), although it is clearly present because $R \rightarrow 1$ in the dc limit.
The low energy $\sigma_2$, the part of optical conductivity representing inductive current, is positive in all four samples, see Fig.~\ref{Fig1:Doping}(d), pointing to a metallic conduction.

The absence of a visible peak in $\sigma_1$ indicates that the scattering rate is extremely small. For example, in sample A the fits of optical reflectivity to Drude-Lorentz model give an upper limit of $1/\tau \lesssim$1~meV ($\sim$10~cm$^{-1}$), which includes the error margin.
It is clear that the polished (001) samples have a comparatively wider Drude contribution than the (112) samples. Moreover, plasma edges are broader and of very similar shapes in the two polished (001) samples, and the Drude-Lorentz analysis gives an estimate of  $1/\tau \sim 6$~meV (50~cm$^{-1}$).
While the (001) surface is optically isotropic, polishing the crystals of Cd$_3$As$_2$ damages the surface by introducing micro cracks. Such a damaged surface then leads to a shorter mean free path for the conduction electrons, which translates into an increased scattering rate, as indeed observed.

It is possible to estimate the effective carrier density in the studied samples through the analysis of the frequency dependent spectral weight. This quantity is the area under the conductivity spectrum up to a chosen cutoff frequency $\omega_c$:
$$W(\omega_c) = \int_{0}^{\omega_c} \sigma_1(\omega) d \omega =\frac{\pi N_{eff}(\omega) e^2}{2m^* V_c},
$$
where $N_{eff}(\omega)$ is the effective number of carriers, $m^*$ is the effective carrier mass, and $V_c$ is the unit cell volume.
The effective carrier number is determined from the spectral weight up to the cutoff photon energy 18~meV (145 cm$^{-1}$), which is exactly below the strong phonon contribution, followed by the onset of the interband absorption.
Spectral weight determined in such a way gives the quantity $N_{eff}/(m^*/m_e)$, which varies from 0.04 electrons per unit cell in sample A, up to 0.2 electrons per unit cell in samples B, C and D (here $m_e$ is the free electron mass). The effective mass is proportional to the Fermi level in a system with linearly dispersive energy bands. From magneto-optical spectroscopy it is known that the effective mass for sample A is $m^*\approx 0.025\, m_e$, compared to $m^*\approx 0.05\, m_e$ in samples B, C and D.\cite{Akrap2016}
Taking into account the effective mass $m^*$, the estimated carrier concentration varies from  $n\approx 2.5\times 10^{17}$~cm$^{-3}$ in sample A, up to $n\approx 2.5 \times10^{18}$~cm$^{-3}$ in samples B, C and D.

Figure \ref{Fig1:Doping}(c) shows the optical conductivity in the quasi-linear regime for photon energies below 0.5~eV. At low temperature, the interband transitions of a three dimensional system with linear energy dispersion are expected to extend to zero photon energy for a vanishingly small carrier density, resulting in $\sigma_1(\omega) \propto \omega$.\cite{TimuskPRB13,Orlita2014} In our samples the carrier density is finite, so the interband absorption sets in at a finite photon energy, $\hbar \omega = E_F$, as illustrated in the inset of Fig.~\ref{Fig1:Doping}(c). The missing interband spectral weight, from $\hbar \omega = 0$ to $\hbar \omega = E_F$, is recovered in the Drude spectral weight.
Quasi-linear conductivity sets in at $\sim 80$ meV in sample A, and at $\sim 200$ meV in samples B and C, indicated by a small cusp in $\sigma_1$.
The interband contribution or the slope of $\sigma_1 (\omega)$ varies little from one sample to the other.
Our previous optical experiments in high magnetic fields imply a low-energy description of the band structure in terms of one heavy-hole band and a large conical band,\cite{Akrap2016} illustrated in the inset of Fig.~\ref{Fig1:Doping}(c). For such a band structure, the onset energy of interband absorption approximately corresponds to the Fermi level, which is then equal to the energy difference between the lowest unoccupied state in the cone and the occupied states in the flat heavy hole band with high density of states.
It is important to note that the quasi-linear slope of $\sigma_1(\omega)$ has a finite intercept at higher energies.\cite{Neubauer2016,TimuskPRB13}

Figures~\ref{Fig1:Doping}(e) and (f) show the real and imaginary part of the dielectric function, $\epsilon_1$ and $\epsilon_2$, determined using ellipsometry at room temperature, measured for the same set of four samples. The experimental data may be compared with recent DFT calculation.\cite{Mosca-Conte2017} Experimentally, the peak positions in $\epsilon_2$ are at 1.5--1.7 eV and 3.5 eV, whereas DFT gives 1.2 eV and 2.9 eV.
While there is good qualitative agreement in the peak positions and the shape of $\epsilon_1$ and $\epsilon_2$, both components of the dielectric function are twice smaller in the experiment than in the calculation.\footnote{It is possible that the spin-polarized calculation in Ref.~\onlinecite{Mosca-Conte2017} counts the states twice.}

There is a striking difference in $\epsilon_1$ and $\epsilon_2$ between the (112) and the (001) oriented samples. The two (001) oriented and polished samples have significantly broadened features. For example, the peak in $\epsilon_2$ at 3.5 eV disappears entirely.
Despite the different dielectric response for different orientations, optical anisotropy at high energies is not necessary to explain our observations. In fact, a polished (112) oriented sample shows a similar $\epsilon_1$ and $\epsilon_2$ to the polished (001) samples, where the peaks in $\epsilon_2$ are smeared out, and the intensity of $\epsilon$ is overall reduced in comparison to samples with as-grown surfaces (Fig. S1 of the Supplementary Materials).
Similar observations have been made in crystalline and amorphous silicon, where sharp features in the dielectric function gradually diminish with increasing microstructural damaging of the samples.\cite{Giri2001}

\begin{figure*}[!t]
\centering
\includegraphics[trim = 0mm 0mm 0mm 0mm, clip=true, width=18cm]{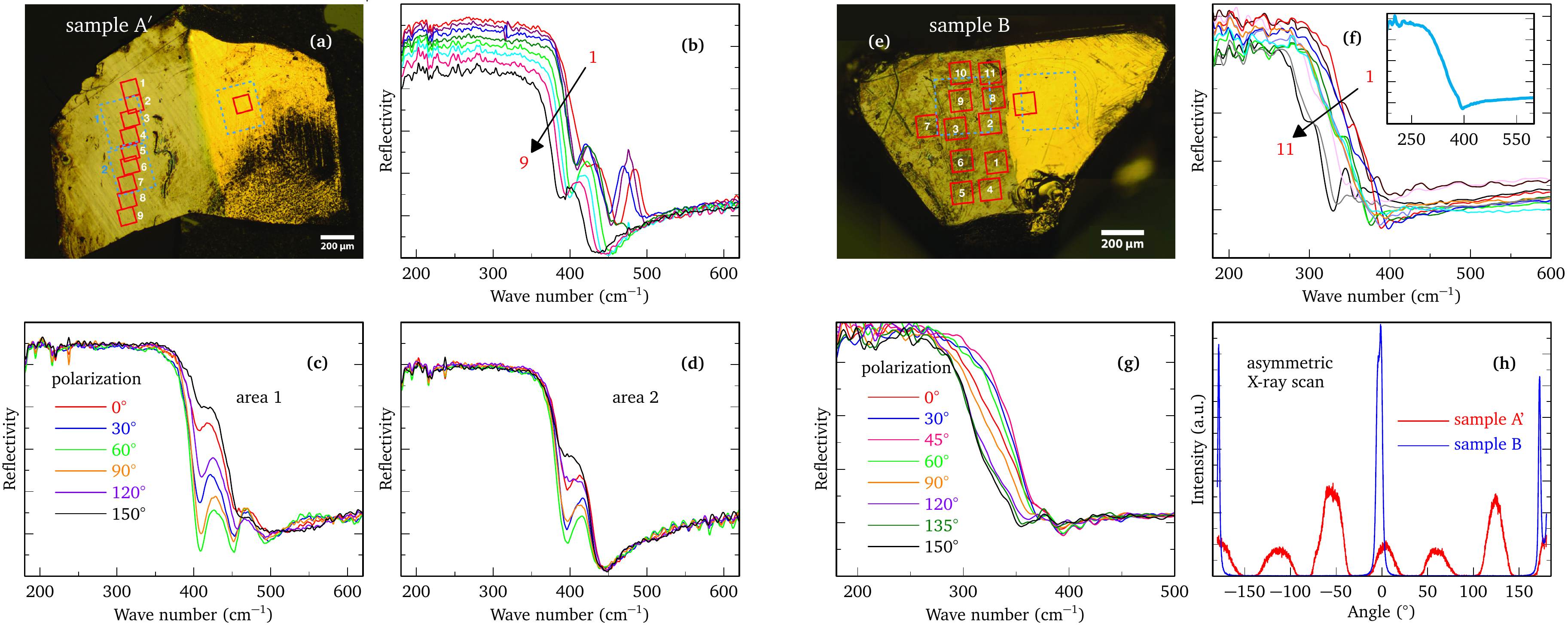}
\caption{\label{Fig3:uscope} (color online) (a) Optical microscope image of sample A$^\prime$. The squares correspond to the probed spots on gold (right) and sample (left) for spatial mapping (red) and polarization dependence (blue dashed). The red squares are approximately 100 by 100 $\mu$m, the blue corresponds to about 250 by 290 $\mu$m. (b) Reflectivity curves of the red spots shown in panel (a). (c), (d) Polarization dependence of the reflectivity of the blue areas (1) and (2) respectively.
(e) Optical microscope image of sample B. The red squares are approximately 100 by 100 $\mu$m, the blue area corresponds to about 260 by 260 $\mu$m. (f)  Reflectivity curves of the red spots shown in panel (e). Inset in (f) shows the reflectivity measured over the larger sample area.
(g) Polarization dependence of the reflectivity of the blue areas in panel (e).
(h) Intensity of the (325) reflection of the asymmetric X-ray scan for samples A$^\prime$ and B.
}
\end{figure*}

Next we focus on the temperature evolution of the optical properties.
Fig.~\ref{Fig2:Temperature} shows the reflectance and the real part of optical conductivity $\sigma_1$ for a dense mesh of temperatures, determined for sample B on its as-grown, lustruous (112) surface. Temperature-dependent spectra on the other three samples are shown in the Fig.~S2 of the Supplementary Materials.

Temperature-dependent reflectance is shown in Fig.~\ref{Fig2:Temperature}(a) in the far infrared region, while the inset zooms in on a mid infrared feature associated with Pauli blocking (Burnstein-Moss effect).\cite{Jenkins2016}
In this particular sample the plasma frequency decreases as the temperature is increased, which is unusual. In a semimetal one would normally expect that there be more carriers as temperature increases. In the three remaining samples (A, C, D) the effect is opposite and the plasma frequency slightly increases as the temperature grows, as seen in Fig.~S2 of the Supplementary Materials.

The plasma edge of sample B develops a step at 55~meV at low temperatures. Such a fine structure of the plasma edge may originate from optical anisotropy, inhomogeneity of charge concentration in bulk, surface charge depletion, or plasmarons as suggested previously.\cite{Jenkins2016}
The step-like structure is translated into a small feature in $\sigma_1$ at 50~meV. Strictly speaking, this feature shows that the validity of Kramers-Kronig approach may be partly limited, as there may be mixing of contributions pertaining to different optical axes. However, the corresponding feature in conductivity is only a small peak at 50~meV, which means that the mixing of these contributions is also small.
A similar step-like feature occurs in all the Cd$_3$As$_2$ samples we have measured, as seen in Fig.~\ref{Fig1:Doping}(a), and will be discussed below in more detail.

A clear onset of interband absorption, or Pauli blocking edge, develops in $\sigma_1$ at low temperature, seen as a step-like feature in $\sigma_1$ at $\sim 200$~meV in Fig.~\ref{Fig2:Temperature}(b). The peak in the reflectance above 200~meV is caused by a logarithmic divergence in $\epsilon_1$  at the onset of interband transitions, as depicted in the inset of Fig.~\ref{Fig1:Doping}(c). The pronounced Pauli edge forms very gradually towards low temperatures, following the sharpening of the Fermi-Dirac distribution. The sharp step reduces the slope of $\sigma_1$ with respect to higher temperatures. While the room temperature slope has a finite extrapolation, the low temperature slope of the interband $\sigma_1(\omega)$ passes approximately through zero.

The temperature dependence of the Drude contribution is shown in Fig.~\ref{Fig2:Temperature}(c). A Drude-Lorentz fit of the data tells that the Drude weight in sample B weakly decreases with increasing temperature, whereas the scattering rate increases with increasing temperature but remains below or close to $10$~cm$^{-1}$ ($\sim 1$~meV).

\subsection{Spatial inhomogeneity of optical response}

Figure \ref{Fig3:uscope}(a) shows a photograph of sample A$'$, originating from the same batch as sample A. The upper angle of the sample reveals the hexagonal symmetry of the exposed (112) plane. The length of the sample is about 2 mm and it is partly covered in gold.
The reflectivity shown in Figs.~\ref{Fig3:uscope}(b)--(d) is determined at 10 K in the infrared microscope, using the gold patch as a reference. The obtained quantity is not an absolute reflectance since there is some tilt to the sample, hence the scale of reflectivity is not absolute. Measurement spots on the sample, defined by aperture blades of the infrared microscope, are indicated by blue and red rectangles. Data taken for the blue area 1 is shown in (c) and for blue area 2 in (d), while red rectangles correspond to data in panel (b). The lateral dimension of red rectangles is approximately 100 $\mu$m and blue rectangles 250 $\mu$m.

Figure~\ref{Fig3:uscope}(b) shows the unpolarized spatial dependence of the optical response. We see that the reflectivity has a pronounced double plasma edge. Overall, we observe strong spatial dependence, yielding even triple plasma edge for some locations.

The complex shape of the plasma edge indicates that there are several different contributions at play. The measured reflectance $R$ is in that case given by a weighed average of different contributions:
$$ R = \sum_i \alpha_i R_i $$
where $R_i$ is the reflectivity of $i$th contribution, and $\alpha_i$ its relative weight.
The reasons for these different contributions include optical anisotropy, crystal twinning, or spatially non-uniform carrier concentration.
For example, a double plasma edge may be caused by the mixing of two different optical axes, when using unpolarized light in an optically anisotropic plane.
The fine structure of the plasma edge may also be due to spatial non-uniformity of the carrier concentration; for example developing of charge puddles, or surface charge depletion. To distinguish between these different possibilities, we study the polarization dependence of the reflectivity, shown in Figs.~\ref{Fig3:uscope}(c) and (d) for two distinct spots. We see that we cannot eliminate the step in the reflectivity for any polarizer angle. In (c), extrema are separated by 90$^{\circ}$. In (d), they are separated by 60$^{\circ}$.
The dependence is markedly different for the two spots, underlining the significance of spatial inhomogeneity in the sample.

To verify whether the sample has single or multiple structural domains, we performed an azimuthal scan in diffraction from the (325) plane.
Diffraction from this plane was selected since it has high scattering factor (high intensity), making it possible to measure in coplanar geometry (with angles of incidence and exit approximately 6$^\circ$ and 24$^\circ$) with no other diffractions nearby.
A single domain sample would have two peaks, 325 and 235 diffractions, for azimuth varied by 180$^\circ$.
Figure~\ref{Fig3:uscope}(h) shows the asymmetric X-ray scan on sample A$^\prime$, which clearly indicates there is crystal twinning.
Three domains are observed, because peaks in the intensity appear at every 60$^{\circ}$. The dominant domain takes up about half the volume, while the remaining two twins cover a smaller volume, roughly one quarter each. Different parts of the sample have different dominant orientation.
These domains differ in orientation of the c-axis along the surface normal, in such a way that the global (112) orientation of the sample is preserved.

\begin{figure*}[t]
\includegraphics[trim = 0mm 0mm 0mm 0mm, clip=true, width=18cm]{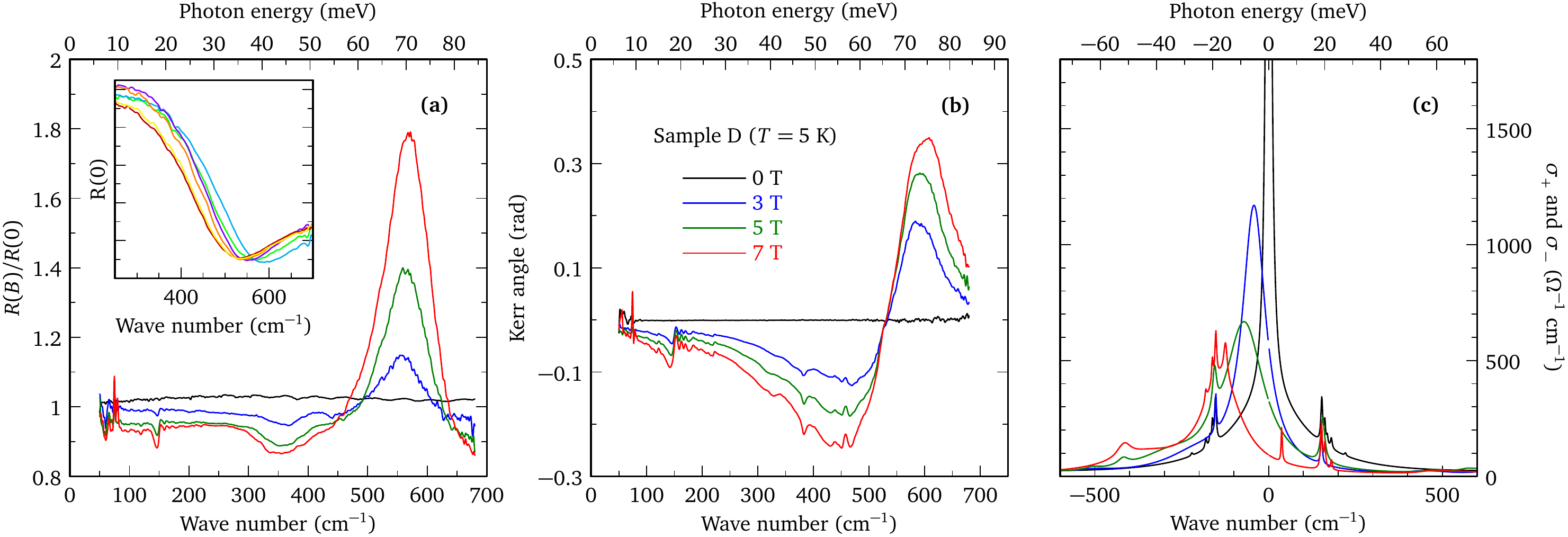}
\caption{\label{Fig3:Kerr} (color online) (a) Relative magneto-reflectivity  and (b)  Kerr angle as a function of photon energy, determined for sample D for several magnetic fields. Inset shows the zero-field spatial variation of the reflectivity, on a 200~$\mu$m scale. (c) Optical conductivity in the circularly polarized basis, $\sigma_+$ and $\sigma_-$, are shown for different magnetic fields. }
\end{figure*}

Figure \ref{Fig3:uscope}(e) shows a photograph of sample B, which like sample A$^\prime$ exposes the (112) plane. In contrast to sample A$^\prime$, sample B contains only a single structural domain. Indeed, the asymmetric X-ray scan in Fig.~\ref{Fig3:uscope}(h) shows only 0$^{\circ}$ and 180$^{\circ}$ peaks for sample B. Surprisingly, even in a single domain sample, the reflectivity strongly varies from one position to another. Figure \ref{Fig3:uscope}(f) shows the reflectivity measured at 11 rectangular spots on this sample. The inset of Fig.~\ref{Fig3:uscope}(f) shows the overall reflectivity of the sample, without strong features at the plasma edge.
Each spot is approximately a square of side 100~$\mu$m. The overall variation in the plasma edge is on the scale of $60-70$~cm$^{-1}$, which can only be explained by the corresponding local variation of the carrier density of approximately 30\%. The same sample had been measured several months prior, as shown in Fig.~\ref{Fig2:Temperature}, and its plasma edge was found to be almost 15~meV higher than in Fig.~\ref{Fig3:uscope}(f) which corresponds to a loss of almost 50\% of the carriers, attributed to cadmium vacancy annealing.

In sample A$^\prime$ we can follow the progression of the plasma edge as we move from one spot to another. There is a distinct redshift when moving from position 1 to 9, as shown in Fig.~\ref{Fig3:uscope}(b). On the contrary, in the single-domain sample B, different plasma edges seem to be randomly scattered on the surface of the sample, see Fig.~\ref{Fig3:uscope}(f). Polarization-dependent spectra show that it is possible to almost completely separate a single contributing plasma edge in sample B, as illustrated in Fig.~\ref{Fig3:uscope}(g). The spread in the plasma edge as a function of polarizer angle is less than 50~cm$^{-1}$, which is smaller than the spatial variation of the plasma edge. Therefore, anisotropy of the optical response alone is not sufficient to produce a strong feature at the plasma edge, especially in presence of more than two charge puddles. If there are no charge puddles, then the anisotropy will produce a sharp feature. However, if there are multiple charge puddles like seen in Fig.~\ref{Fig3:uscope}(f), they will broaden the plasma edge (see Fig.~S4).

Taken together, these measurements show that the charge density changes locally in Cd$_3$As$_2$, and this non-uniformity of charge density is not necessarily linked to the crystal twinning. Whether the samples are single domain or not, the spatial dependence of plasma frequency is present in all the samples we have investigated.
When different structural domains are present, they appear to reinforce the carrier density variation.

The spatial scale of the charge puddles is on the order of 100~$\mu$m. If their size were significantly smaller, the contributions would mix and one would not observe a significant difference when moving from one spot to another.
Charge puddles in Cd$_3$As$_2$ may be linked to the annealing at room temperature that takes place over a long period of time after the sample synthesis. Cadmium atoms progressively order within the tetrahedral voids of the crystal structure, and this atomic diffusion happens on a time scale of months or even years.\cite{Arushanov1986aa}
Indeed, in all the samples investigated here, the plasma edge shifted to lower frequencies with time (see Fig.~S3 of Supplementary Materials). This shift translates into a decreased carrier density over time. There is a priori no reason why annealing should be spatially uniform, and it may reinforce the charge puddles.

When several charge puddles are present in a sample, such as Figs.~\ref{Fig3:uscope}(b) and (f) show, the total reflectivity should have a broadened and smooth aspect (See Section IV of Supplementary Materials).
In contrast, a sharp feature in reflectance is typically observed in the (112) samples and shown in Figs.~\ref{Fig1:Doping}(a) and \ref{Fig2:Temperature}(a). Such a sharp feature at the plasma edge can only arise when two dominant contributions are in play.
This implies that the sharp feature in the (112) oriented samples may be linked to optical anisotropy, where different optical axes have different plasma frequencies, but only when there are no more than two charge puddles present. Alternatively, a sharp feature at the plasma edge can be produced in a twinned crystal structure where each twin has a different charge density.

In a three-dimensional cone the carrier density goes as $n\propto E_F^3$. This means that for a 30\% change in local carrier density, one can expect a 12\% change in the local value of Fermi energy. In addition, spatial variation of the carrier density in Cd$_3$As$_2$ may lead to some variation of mobility $\mu$, for instance through the effective mass, $\mu \propto \tau/m^*$.

\subsection{Magneto-optical Kerr effect}

When a sample is placed in a magnetic field, the reflectivity plasma edge splits into two branches, historically known as cyclotron resonance inactive and active modes. These branches are separated in energy by the value of cyclotron resonance energy, $\hbar\omega_c$.\cite{PalikRPP70}
Therefore, strong magnetic field-dependent features are expected to appear surrounding the zero field screened plasma edge, both in the relative magneto-reflecticity and Kerr rotation measurements. Figs.~\ref{Fig3:Kerr}(a) and (b) show the  relative magneto-reflectivity and Kerr effect determined for the (001)-oriented single-domain sample D in magnetic fields up to 7~T.\footnote{Our experimental method to measure Kerr angle is limited to optically isotropic planes, and in the case of orthogonal Cd$_3$As$_2$ this is the (001) plane.}
The plasma edge splitting in $R(B)/R(0)$ is observable as a dip--hump structure around the screened plasma frequency, $\omega_p$. At energies below $\omega_p$ the reflectivity is reduced by magnetic field, whereas above $\omega_p$ it is strongly enhanced. However, the experimental curves cannot be reduced to a single dip--hump structure, but instead show additional features. For example, there are apparent oscillations in the range between 350 and 500~cm$^{-1}$.
These weaker features are fully consistent with the charge inhomogeneity observed in the microscopy experiments in zero field. The spatial distribution of plasma edges of sample D is shown in the inset of Fig.~\ref{Fig3:Kerr}(a) and implies a 30\% spatial variation of the carrier density.
Sharp features in $R(B)/R(0)$ close to 150~cm$^{-1}$ are due to phonons. As magnetic field increases, the Drude peak at $\omega=0$ is replaced by a cyclotron resonance at a finite frequency. The phonons are thereby screened less, and appear more prominently in magnetic field.

The Kerr angle measures how much the polarization state of a linearly-polarized incident beam of light rotates upon reflecting off a sample in a magnetic field. The Kerr angle in Cd$_3$As$_2$ is very large, with maximum rotation polarization reaching 0.35~rad (20$^\circ$) at 7~T. For a single split plasma edge the Kerr rotation has an antisymmetric shape in frequency, where the zero angle crossing corresponds to the screened plasma frequency $\omega_p$ at zero field, and the slope of the resonance reveals the sign of the charge carriers. In the case of Cd$_3$As$_2$ the positive slope indicates electron-like charge carriers. As evidenced by the more complex shape of the resonance in Fig.~\ref{Fig3:Kerr}(b), notably the additional kinks in the slope around $\omega_p$, the Kerr angle measurements further underline the influence of the inhomogeneous carrier density.

For further analysis of the magneto-optical spectra and extraction of the complex conductivity tensor, we use the magneto-optical Kramers-Kronig relations.\cite{Levallois2015} The magneto-reflectivity gives access to the diagonal elements of the optical tensor describing the dielectric response, $\tilde{\epsilon}$ or $\tilde{\sigma}$. Because of the external field, the optical response tensor also acquires off-diagonal elements, which are proportional to the Kerr angle:
$$ 	\tilde\epsilon(\omega) =
	\begin{bmatrix}
	\epsilon_{xx} & \epsilon_{xy}\\
	\epsilon_{yx} & \epsilon_{xx}	\\
	\end{bmatrix} $$
The response tensor can be diagonalised in the basis of circularly polarized light with eigenvalues $\epsilon_{\pm}(\omega) = \epsilon_{xx}(\omega)\pm i \epsilon_{xy}(\omega)$. These eigenvalues,  $\epsilon_{+}$ and $\epsilon_{-}$, are related to the optical conductivity via the relation: $\epsilon_{\pm}(\omega) = 1+ 4\pi i \sigma_{\pm}(\omega)/\omega$. From Fresnel equations, the experimentally obtained reflectivity and Kerr angle are related to the complex magneto-optical response tensor $\epsilon_{\pm}$.
The experimental data is modeled by a sum of magneto-optical Drude--Lorentz oscillators:
$$ \epsilon_{\pm}(\omega) = \epsilon_{\infty \pm}+\sum_{k}\frac{\omega_{p,k}^{2}}{\omega_{0,k}^{2}-\omega^{2}-i\gamma_{k}\omega\mp\omega_{c,k}\omega}$$
The real parts of right-handed (positive frequency axis) and left-handed (negative frequency axis) circular conductivity can directly be inferred from the Drude--Lorentz modeling of the data and are shown in Fig.~\ref{Fig3:Kerr}(c). Excitations observed in $\sigma_{-}(\omega)$ correspond to electron-like cyclotron resonance, which due to a finite scattering rate has a tail in the right handed circular conductivity, $\sigma_{+}(\omega)$. Indeed, we see a strong cyclotron resonance with a $1/\tau \sim 7$~meV. This corresponds well to the scattering rate obtained from the zero field conductivity, approximately 6~meV for this sample. An additional wider and weaker resonance appears at higher frequencies.
The existence of charge puddles leads to a spatial dependence of the plasma edge. This in turn strongly influences the magneto-reflectivity and Kerr rotation measurements, and so the experimental results cannot be modelled by a single, or even a double cyclotron resonance. The observation of an artificially wide additional cyclotron resonance is consistent with charge puddles, that is, it mimics a distribution of carrier concentration over the sample's surface. Phonons are observed equally in left and right handed conductivity. From the peak position of the cyclotron resonance, $\omega_c$ = 15 meV at 7 T, we can calculate the effective mass,
$ m^*=\frac{eB}{\omega_c}\approx 0.05$~$m_e,$
where m$_e$ is the free electron mass. For comparison, the effective mass is 0.025 and 0.04~$m_e$ in samples A and C.\cite{Akrap2016}
The carrier mobility $\mu$ is inversely proportional to the width $\gamma$ of the cyclotron resonance, $\mu = \omega_c/(\gamma B)$. However, in our case it is not possible to determine the mobility as the cyclotron resonance peak is spread out as a result of charge puddles.

It cannot be excluded that some of the signatures observed in magneto-optical spectra originate from a surface layer with a shifted chemical potential, like a depletion layer.
Taking a quasi-classical view of bands, one may consider a depleted layer on the surface. Smaller density of carriers would produce a smaller effective mass, explaining why a broad, weak resonance is found at a higher photon energy. In this picture the Fermi level on the surface would be 2.5 times smaller when compared to bulk. The broadness of this cyclotron resonance would imply some distribution of Fermi energies.

\section{Conclusions}

We have studied semimetallic Cd$_3$As$_2$ by means of optical spectroscopy.
All of the investigated samples show a relatively small carrier density of the order $10^{17}-10^{18}$~cm$^{-3}$ and a very small Drude scattering rate ($\sim 1$~meV), which is increased in the polished samples.
In all samples, carrier density is nonuniform and it varies by approximately 30\% on the scale of $\sim 100$~$\mu$m across the surface of the crystal.
Optical conductivity has a quasi-linear dependence on photon energy in a wide range.
Temperature dependent spectra show a clear development of the Pauli blocking edge, where the interband absorption sets in. A small optical anisotropy in the (112) plane becomes evident only at low temperatures. Kerr rotation and magneto-reflectivity provide us with the magneto-optical conductivity in the circular basis, which confirms that the charge carriers are electrons, but indicates there is more than one contribution to the conductivity, consistent with charge puddles in Cd$_3$As$_2$.

The local variation of optical properties of Cd$_3$As$_2$ is increased by the crystal twinning which is observed in some of the samples. The low-energy properties of Cd$_3$As$_2$ are found to vary with time, most likely due to the annealing of cadmium vacancies, and the development of charge puddles.

Let us note that the charge puddles, whose presence was demonstrated in all samples explored by this work, are most likely a general characteristics of Cd$_3$As$_2$ crystals. The fine structure of the plasma edge, which serves as a good indication of such puddles, was also observed in optical studies performed on samples from a completely different source.\cite{SchleijpeIJIMW83} The probable appearance of areas with distinctively different electron densities should therefore be taken into account, for example when analysing data obtained in quantum oscillations experiments (Shubnikov-de Haas or de Haas-van Alphen), which often report double or even multiple frequency oscillations in Cd$_3$As$_2$.

\section{Acknowledgements}
I.C. acknowledges funding from the Postdoc.Mobility fellowship of the Swiss National Science Foundation.
A.A. acknowledges funding from the Ambizione Fellowship of the Swiss National Science Foundation.
This work has been supported by the ERC via project MOMB and by MEYS project CEITEC 2020 (No. LQ1601), as well as ANR DIRAC3D. Work at BNL was supported by the U.S. Department of Energy, Office of Basic Energy Sciences, Division of Materials Sciences and Engineering under Contract No. DE-SC0012704.
Part of this work was done at Soleil, proposal number 20151043 on SMIS beamline.
Work at the Institute for Quantum Matter (IQM) was supported by the U.S. Department of Energy, Office of Basic Energy Sciences, Division of Materials Sciences and Engineering through Grant No. DE-FG02-08ER46544. T.M.M. acknowledges support of the David and Lucile Packard Foundation. Partial funding for this work was provided by the Johns Hopkins University Catalyst Fund.

The authors acknowledge illuminating discussions with Krzystof Grasza, Alexey Kuzmenko and Neven Bari\v{s}i\'c, generous help with the samples from N. Peter Armitage and Frederic Teppe, as well as helpful comments from Nathaniel Miller.

\bibliography{Cd3As2}

\newpage
\pagenumbering{gobble}

\begin{figure}[htp]
\includegraphics[page=1,trim = 17mm 17mm 17mm 17mm, width=1.0\textwidth,height=1.0\textheight]{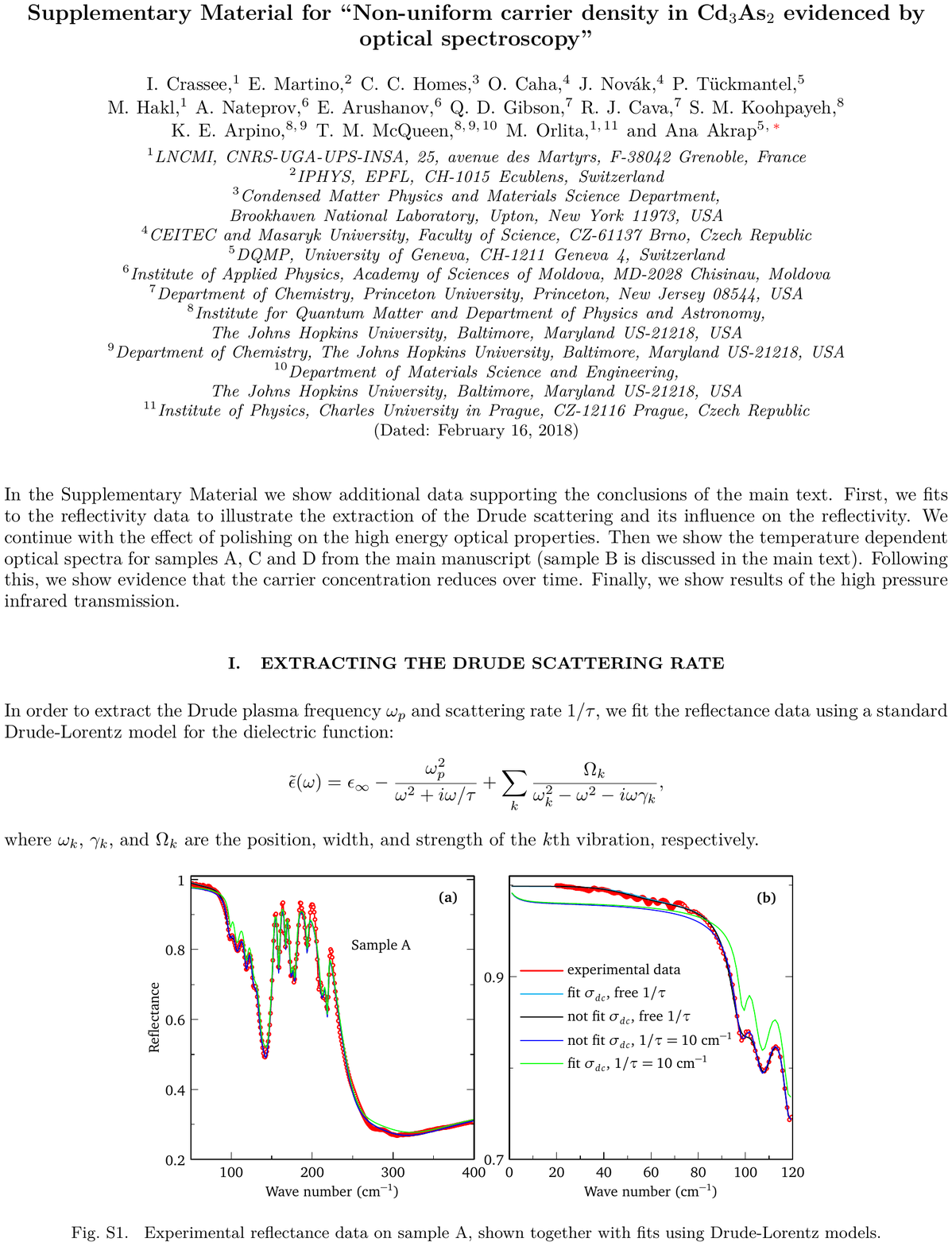}

\end{figure}

\newpage

\begin{figure}[htp]
  \includegraphics[page=2,trim = 17mm 17mm 17mm 17mm, width=1.0\textwidth,height=1.0\textheight]{SI.pdf}

\end{figure}

\newpage

\begin{figure}[htp]
  \includegraphics[page=3,trim = 17mm 17mm 17mm 17mm, width=1.0\textwidth,height=1.0\textheight]{SI.pdf}

\end{figure}

\newpage

\begin{figure}[htp]
  \includegraphics[page=4,trim = 17mm 17mm 17mm 17mm, width=1.0\textwidth,height=1.0\textheight]{SI.pdf}

\end{figure}

\begin{figure}[htp]
  \includegraphics[page=5,trim = 17mm 17mm 17mm 17mm, width=1.0\textwidth,height=1.0\textheight]{SI.pdf}

\end{figure}

\end{document}